\RequirePackage{fix-cm}
\documentclass[smallextended]{svjour3}       % onecolumn (second format)
\smartqed  % flush right qed marks, e.g. at end of proof
\usepackage{graphicx}
\usepackage{amsfonts}
\usepackage{amsmath}
\usepackage{amssymb}
\usepackage{enumerate}
\journalname{General Relativity and Gravitation}
\begin{document}

\title{Emergent Gravity as the Eraser of Anomalous Gauge Boson Masses, and QFT-GR Concord}
%}
%\subtitle{Do you have a subtitle?\\ If so, write it here}

\titlerunning{Emergent Gravity}        % if too long for running head

\author{Durmu{\c s}  Demir}

%\authorrunning{Short form of author list} % if too long for running head

\institute{Durmu{\c s} Demir \at Faculty of Engineering and Natural Sciences, Sabanc{\i} University, 34956 Tuzla, {\.I}stanbul, Turkey\\
                          Webpage: http://myweb.sabanciuniv.edu/durmusdemir/\\
                          \email{durmus.demir@sabanciuniv.edu}}

\date{Received: date / Accepted: date}
% The correct dates will be entered by the editor

\maketitle

\begin{abstract}
In the same base setup as Sakharov's induced gravity, we investigate emergence of gravity in effective quantum field theories (QFT), with particular emphasis on the gauge sector in which gauge bosons acquire anomalous masses in proportion to the ultraviolet cutoff $\Lambda_\wp$. Drawing on the fact that $\Lambda_\wp^2$ corrections explicitly break the gauge and Poincare symmetries, we find that it is possible to map $\Lambda_\wp^2$ to spacetime curvature as a covariance relation and we find also that this map erases the anomalous gauge boson masses. The resulting framework describes gravity by the general relativity (GR) and matter by the QFT itself with $\log\Lambda_\wp$ corrections (dimensional regularization). This QFT-GR concord predicts existence of new physics beyond the Standard Model such that the new physics can be a weakly-interacting or even a non-interacting sector comprising the dark matter, dark energy and possibly more. The concord has consequential implications for collider, astrophysical and cosmological phenomena.
% \PACS{PACS code1 \and PACS code2 \and more}
% \subclass{MSC code1 \and MSC code2 \and more}
\end{abstract}

\section{Introduction}
\label{sect:intro}
The problem of reconciling QFTs with the GR has been under intense study for several decades. The concord  between the special relativity and quantum mechanics that led to QFTs seems to be not evident, if not absent, when it comes to the GR and QFTs. This problem has been approached by exploring various possibilities.  

The first possibility is to quantize the GR but it is known to be not a renormalizable field theory \cite{thooft,stelle}. In general, quantization of gravity is not a straightforward task, and this difficulty has given cause to various alternative approaches \cite{qgr-review} (string/M theory, loop quantum gravity, asymptotic safety, supergravity, $\dots$). The problem here is actually absence of a renormalizable ultraviolet (UV) completion of the GR (like completion of the Fermi theory by 
the electroweak theory).  Even so, it must be kept in mind that, independent of the details of the UV completion, leading quantum corrections 
can be reliably calculated as they are determined by the low-energy couplings of the massless fields in the completion \cite{donoghue,holdom}.  By a similar token, asymptotic safety of the GR is known to lead to a reliable prediction of the Higgs boson mass \cite{safety}. 

The second possibility is to keep the GR classical and make QFT to fit to it \cite{qft-cuvred}. This route, which may be motivated by the views \cite{qgr2} that the gravity could be fundamentally classical, is hindered by the fact that curved spacetime does not allow for preferred states like the vacuum and detectable structures like the particles (positive frequency Fourier components) \cite{cgr2}.  Indeed, allowing no privileged coordinates, curved spacetimes do not accommodate a unique vacuum state as diffeomorphisms shuffle negative and positive frequency modes and one observer's vacuum can become another observer's excited state. Besides,  incoming and outgoing particles and their scattering amplitudes are properly defined mainly in symmetric \cite{woodard} and asymptotically flat \cite{hawking} spacetimes. These hindering features of the curved geometry have been attempted to overcome, respectively, by promoting the operator product expansion in QFTs to a fundamental status \cite{wald-ilk} and by introducing microlocal spectrum conditions \cite{fredenhagen}. These attempts, which leave the mechanism underlying the Newton-Cavendish constant and other couplings to a future quantum gravity theory,  have led to axiomatic QFTs \cite{wald-sonra} in curved spacetime. 

The third possibility is to induce the GR from fluctuations of the quantum fields in flat spacetime. (It is possible to consider also curved spacetime but for loop-induced GR to arise outrightly one must find a way of removing at least the Cavendish-Newton constant in the classical action, let alone the problems noted in \cite{wald-ilk}. One possibility is to invoke classical scale invariance but it removes  not only the Cavendish-Newton constant but also all the field masses, including the Higgs mass \cite{scale}.) This approach, which rests on flat spacetime where QFTs work properly with well-defined vacuum states and particle spectra \cite{wald-ilk,wald,ashtekar}, has the potential to avoid principal difficulties with curved spacetime QFTs \cite{wald,fredenhagen}. To this end, Sakharov's induced gravity \cite{sakharov} (see also \cite{ruzma}, and see especially the furthering analysis in \cite{visser}) provides a viable route as it starts with flat spacetime QFT, and induces curvature sector at one loop through the curved metric for the fields in the loops. The gravitational constant, viewed as metrical elastic constant of the spacetime \cite{sakharov,ruzma,demir2015}, emerges in a form  proportional to the UV cutoff \cite{sakharov,visser,demir2017}, with implications for black holes \cite{fursaev} and Liouville gravity \cite{liouville}. (Besides the Sakharov's, various attempts have been made to generate gravity from quantized matter such as entropic gravity \cite{emerge-verlinde}, entanglement gravity \cite{emerge-raamsdonk}, analog gravity \cite{emerge-analog}, broken symmetry-induced gravity \cite{adler}, Nambu-Jona-Lasinio type gravity \cite{terazawa}, and agravity \cite{strumia}, with a critical review in \cite{emerge-lit}).

The present work has the same base setup as Sakharov's induced gravity: A flat spacetime effective QFT with the usual $\log\Lambda_\wp$, $\Lambda_\wp^2$ and $\Lambda_\wp^4$
loop corrections, $\Lambda_\wp$ being the UV cutoff \cite{sakharov,visser}. The loop corrections, giving cause to various UV sensitivity problems \cite{veltman,ccp,ccb},
are specific to the flat spacetime \cite{demir2019} such that  $\Lambda_\wp^2$ breaks explicitly both the gauge and Poincare (translation) 
symmetries \cite{wigner} while $\log\Lambda_\wp$ breaks none of them. This symmetry structure paves the way for a covariance relation between $\Lambda_\wp^2$ and (Poincare-breaking)
spacetime curvature, with intact $\log\Lambda_\wp$. This novel covariance relation retains the effective QFT with sole $\log\Lambda_\wp$ corrections (dimensional regularization) \cite{dim-reg2,dim-reg3}, and 
induces the GR in a way removing the anomalous gauge boson masses (which arise from matter loops in exact proportion to $\Lambda_\wp$ and which explicitly break charge and color (CCB) symmetries \cite{ccb}).  The end result is a QFT-GR chime or concord. 

In what follows, Sec. \ref{sect:problems} explicates the UV sensitivity problems, including the explicit CCB stemming from anomalous gauge boson masses \cite{anomaly,ccb,demir2016}.  Sec. \ref{sect:hownot} reveals that effective QFTs,  unlike classical field theories,  necessitate, in getting to curved spacetime, curvature sector to be spanned exclusively by the QFT mass scales. Sec. \ref{sect:constantlog} shows that $\Lambda_\wp^2$ breaks explicitly both the gauge and Poincare symmetries while $\log\Lambda_\wp$ respects both of them, and one takes note of the Poincare affinity between $\Lambda_\wp^2$-induced broken Poincare (translation) invariance and nonzero spacetime curvature, with intact $\log\Lambda_\wp$. 
Sec. \ref{sect:restore} extends the usual general covariance between the flat and curved metrics by introducing a covariance relation between $\Lambda_\wp^2$  and affine curvature, and shows by a step-by-step analysis that 
the anomalous gauge boson masses are wiped out 
(namely, CCB is solved dynamically up to doubly Planck-suppressed terms) by the emergence of curvature. Sec. \ref{sect:concord} carries the entire flat spacetime effective QFT into curved spacetime via the extended general covariance, and shows that an intertwined whole of a purely loop-induced GR plus the same QFT with sheer $\log\Lambda_\wp$ regularization (equivalent to dimensional regularization \cite{cutoff-dimreg}) emerges to form a QFT-GR concord.  Sec. \ref{sect:concord} discusses some
salient aspects of the concord in view of the Starobinsky inflation \cite{starobinsky,Planck}, cosmological constant problem (CCP) \cite{ccp,ccp2}, little hierarchy problem \cite{little}, and induction and field-dependence of the  Newton-Cavendish constant \cite{non-min,codata}. Sec. \ref{sect:concord} also 
discusses the standard model (SM) and shows how symmergence necessitates a new physics sector \cite{demir2016,demir2019} and what effects this new physics sector might have on 
cosmological \cite{de,darksector,cos-coll}, astrophysical \cite{cdm,darksector} and  collider \cite{collider1,collider2} phenomena.  Sec. \ref{sect:conc} concludes the 
work, and gives prospects for future research by highlighting the salient aspects of the QFT-GR concord.  

\section{UV Cutoff and the UV Sensitivity Problems}
\label{sect:problems}
The UV cutoff $\Lambda_\wp$ is a physical scale. It cuts off the loop momenta to lead to finite physical loop corrections.  It is not subject to any
specific bound (like unitarity bound from graviton exchange) in flat spacetime. Indeed, for flat spacetime QFTs gravity is a completely alien interaction. 
For incorporating gravity into QFTs,  QFTs must be endowed with a mass scale evoking  the fundamental scale of gravity or the curvature.
The UV cutoff $\Lambda_\wp$  is  the aforementioned mass scale. It has a physical correspondent in the gravity sector. It therefore is not a regularization scale. 
Its effects cannot therefore be imitated by the cutoff regularization \cite{polchinski} or the dimensional regularization \cite{dim-reg2,dim-reg3} or any other regularization scheme.

In the presence of the  UV momentum cutoff $\Lambda_\wp$, each coupling in a flat spacetime QFT  (of various fields $\psi_i$ with masses $m_i$) develops a certain $\Lambda_\wp$ sensitivity via the matter loops \cite{eff-qft,polchinski,weinberg-eff}. The sensitivity varies with the mass dimension of the 
operators involved and, in this regard,  a systematic analysis of particle masses and vacuum energy proves useful:
\begin{enumerate}[(1)]
\item Massless gauge bosons $V_\mu$ (like the photon and the gluon) acquire masses 
\begin{eqnarray}
\label{deltMV}
\delta M_V^2 = c_V \Lambda_\wp^2
\end{eqnarray}   
with a loop factor $c_V$ that can involve $\log \Lambda_\wp$ at higher loops. These purely loop-induced masses are plain quadratic in $\Lambda_\wp$, that is, there exist no logarithmic contributions to $\delta M_V^2$ (as they live in the transverse part of the $V_\mu$ self-energy). It is clear that $\delta M_V^2$ are anomalous gauge boson masses, which give cause to explicit CCB \cite{ccb} (see \cite{casas} for spontaneous CCB), as exemplified in Table \ref{table-1} for the SM gauge bosons. Here, it must be emphasized that $\delta M_V^2$ are physical corrections rather than cutoff regularization terms, and their effects therefore are not subject to any (finite or otherwise) renormalization prescription, and the anomalies they give cause to are true anomalies.

\item The corrections to scalar masses 
\begin{eqnarray}
\label{deltmphi}
\delta m_\phi^2 = c_\phi \Lambda_\wp^2 + \sum_i {c}^{(l)}_{ \phi \psi_i} m_i^2 \log \frac{m_i^2}{\Lambda_\wp^2} 
\end{eqnarray}
contain both $\Lambda_\wp^2$ and $m_i^2$ contributions. The former, whose loop factor $c_\phi$ is given in Table \ref{table-1} for the SM Higgs boson, 
gives rise to the big hierarchy problem \cite{veltman}.  The logarithmic $m_i^2$ contribution, on the other hand,  involves
a loop factor ${c}^{(l)}_{ \phi \psi_i}$, and gives cause for the little hierarchy problem \cite{little}. 

\item Finally, the shift in the vacuum energy
\begin{eqnarray}
\label{deltV}
\delta V = c_\varnothing \Lambda_\wp^4 + \sum_i {c}_{\psi_i} m_i^2 \Lambda_\wp^2 +  \sum_i {c}^{(l)}_{\varnothing \psi_i} m_i^4 \log \frac{m_i^2}{\Lambda_\wp^2} 
\end{eqnarray}
involves quartics and quadratics of both $\Lambda_\wp$ and $m_i$. The loop factors $c_\varnothing$ and ${c}_{\psi_i}$ are given 
in Table \ref{table-1} for a generic QFT as well as the SM. The shift in the vacuum energy gathers both 
scales marginally. It  causes  no problem (like the CCP \cite{ccp,ccp2}) in flat spacetime.
\end{enumerate}

\begin{table}
\caption{\label{table-1} One-loop corrections in a generic QFT (first column) and 
in the SM (other columns).  Here, 
${\rm str}[1] = \sum_{s} (-1)^{2s} (2 s + 1) {\rm tr}[1_s] = n_b - n_f$,
where $n_{b(f)}$ is the number of bosons (fermions) and  ${\rm tr}[1_s]$ traces over 
charges of fields with spin $s$ (like electric and color charges). Likewise, ${\rm str}[m^2] = 
\sum_{s} (-1)^{2s} (2 s + 1) {\rm tr}[m_s^2]$. The top quark mass is denoted by $m_t$,  
$W$ boson mass by $M_W$,   strong coupling by $g_s$, 
weak coupling by $g_2$, and the hypercharge gauge coupling by $g_Y$.}
\begin{tabular}{|l|l|l|l|}
\hline
QFT Value&SM Fields&SM Value &Problem Caused in SM \\
\hline\hline
$c_V$ & gluon  & $\frac{21 g_s^2}{16 \pi^2}$ & color breaking\\\hline
$c_V$ & weak boson &$\frac{21 g_2^2}{16 \pi^2}$ & isospin breaking\\\hline
$c_V$ & hypercharge boson  &$\frac{39 g_Y^2}{32 \pi^2} $ & hypercharge breaking\\\hline
$c_\phi$ & Higgs boson &$\frac{g_2^2 {\rm str}[m^2]}{8 \pi^2 M_W^2}\approx - \frac{g_2^2 m_t^2}{\pi^2 M_W^2}$ & big hierarchy problem\\\hline
$c_\varnothing = -\frac{{\rm str}[1]}{128 \pi^2}$& over all SM fields & $ \frac{31}{32 \pi^2}$ &no CCP (flat spacetime)\\\hline
$\sum_{i} c_{\psi_i} m_i^2=\frac{{\rm str}[m^2]}{32 \pi^2}$ & over all SM fields & $\approx -\frac{m_t^2}{4 \pi^2}$ &no CCP (flat spacetime)\\
\hline
\end{tabular}
\end{table}

\section{How Not to Take Effective QFTs into Curved Spacetime}
\label{sect:hownot}
Classical field theories \cite{landau}, 
governed by actions  $S_{c}\left(\eta,\psi,\partial\psi\right)$ of various fields $\psi_i$ in the flat spacetime of 
metric $\eta_{\mu\nu}$,  are carried into curved spacetime of a metric $g_{\mu\nu}$ by letting
\begin{eqnarray}
S_{c}\left(\eta,\psi,\partial\psi\right) \hookrightarrow S_{c}\left(g,\psi,\nabla\psi\right) + ``{\rm curvature\ sector}"
\label{map-ac}
\end{eqnarray}
in accordance with general covariance \cite{norton}, which itself is expressed by the map
\begin{eqnarray}
\label{covariance-map}
\eta_{\mu\nu} \hookrightarrow g_{\mu\nu}\,,\; \partial_\mu \hookrightarrow \nabla_\mu
\end{eqnarray}
such that the Levi-Civita connection
\begin{eqnarray}
{}^g\Gamma^\lambda_{\mu\nu} = \frac{1}{2} g^{\lambda\rho}\left(\partial_{\mu} g_{\nu\rho} + \partial_{\nu} g_{\rho\mu} - \partial_{\rho} g_{\mu\nu}   \right)
\label{LC-conn}
\end{eqnarray}
sets the covariant derivative $\nabla_\mu$ in (\ref{covariance-map}), and gives rise to the Ricci curvature 
\begin{eqnarray}
R_{\mu\nu}({}^g\Gamma) = \partial_\alpha {}^g\Gamma^\alpha_{\mu\nu} - \partial_\nu {}^g\Gamma^\alpha_{\alpha \mu} +  {}^g\Gamma^\beta_{\mu\nu}  {}^g\Gamma^\alpha_{\alpha\beta} - {}^g\Gamma^\beta_{\mu\alpha}  {}^g\Gamma^\alpha_{\nu\beta}
\label{Ricci-met}
\end{eqnarray} 
as well as the scalar curvature $R(g)=g^{\mu\nu}R_{\mu\nu}({}^g\Gamma)$. The curvature sector in (\ref{map-ac}), added by hand for the curved metric $g_{\mu\nu}$ to be able to gain dynamics, must be of the specific form
\begin{eqnarray}
\!\!\!``{\rm curvature\ sector}" \! = \! \int\! d^4x \sqrt{-g}\!\left\{-\frac{R(g)}{16 \pi {\widetilde{G}}} - {\tilde{c}}_2 (R(g))^2 - {\tilde{c}}_3 {\widetilde{G}}(R(g))^3 + \dots \right\}
\label{curv-sector-0}
\end{eqnarray}
for it to be able reproduce the GR at the leading order (with gravitational constant ${\widetilde{G}}$ and curvature couplings ${\tilde{c}}_i$). The unknown constants in (\ref{curv-sector-0}), which are at the same footing as the bare parameters in the QFT sector, cannot in general be avoided or limited simply because curvature sector (kinetic terms for $g_{\mu\nu}$) can contain arbitrary curvature invariants \cite{kretschmann} (in contrast to gauge theories where renormalizability singles out a specific kinetic term). 

The success with classical field theories prompts an important question: Can general covariance carry also effective QFTs into curved spacetime? This
question is important because  effective QFTs, obtained by integrating out all high-frequency quantum fluctuations of the QFT fields, are reminiscent of the
classical field theories in view of their long-wavelength field spectrum (with loop-corrected tree-level couplings).  The answer lies in the effective action (leaving out higher-dimension operators) \cite{demir2019,demir2017}
\begin{eqnarray}
\label{eff-ac}
\!\!S_{eff}\left(\eta,\psi,\Lambda_\wp\right) \!=\! S_{c}\left(\eta,\psi\right) +   \delta S_{l}\left(\eta,\psi,\log\Lambda_\wp\right) + \delta S_{\varnothing\phi}\!\left(\eta,\Lambda^2_\wp\right) + \delta S_V\left(\eta,\Lambda^2_\wp\right)
\end{eqnarray}
in which $S_{c}\left(\eta,\psi\right)$ is the classical action,
\begin{eqnarray}
\delta S_l(\eta,\psi,\log\Lambda_\wp) \!\supset\!\!
\sum\limits_{i} \!\!\int\!\! d^4x \sqrt{-\eta}\! \left\{\!-{c}_{\varnothing\psi_i} m_i^4 \log \frac{m_{i}^2}{\Lambda_\wp^2} - {c}_{ \phi \psi_i} m_i^2  \log \frac{m_{i}^2}{\Lambda_\wp^2} \phi^\dagger \phi \right\}
\label{deltSlog}
\end{eqnarray}
is the logarithmic action composed of the $\log\Lambda_\wp$ parts of (\ref{deltV}) and (\ref{deltmphi}), 
\begin{eqnarray}
\delta S_{\varnothing\phi} \left(\eta,\Lambda^2_\wp\right) &= \int d^4x \sqrt{-\eta} \left\{-c_\varnothing \Lambda_\wp^4 -  \sum\limits_{i} c_{\psi_i} m_i^2 \Lambda_\wp^2 -  c_\phi  \phi^{\dagger} \phi \Lambda_\wp^2\right\}\label{deltSOphi}
\end{eqnarray}
is the vacuum plus scalar mass action gathering $\Lambda_\wp^4$ and $\Lambda_\wp^2$ parts of (\ref{deltV}) and (\ref{deltmphi}), and 
\begin{eqnarray}
\delta S_V\left(\eta,\Lambda^2_\wp\right) &= \int d^4x \sqrt{-\eta} {c_V} \Lambda_\wp^2 {\mbox{tr}}\!\left[\eta_{\mu\nu} V^{\mu} V^{\nu}\right] \label{deltSV}
\end{eqnarray}
is the anomalous gauge boson mass action formed by the purely quadratic corrections in (\ref{deltMV}). 

If the effective QFTs described by (\ref{eff-ac}) are really like the classical field theories then the 
general covariance map in (\ref{covariance-map}) take them into curved spacetime as
\begin{eqnarray}
S_{eff}\left(\eta,\psi,\Lambda_\wp\right) \hookrightarrow  S_{eff}\left(g,\psi,\Lambda_\wp\right) + ``{\rm curvature\ sector}"
\label{map-eff-ac}
\end{eqnarray}
in parallel with the transformation of the classical action in (\ref{map-ac}), with the curvature sector defined in (\ref{curv-sector-0}). The problem with this transformation is that the parameters ${\widetilde{G}}$, ${\tilde{c}_2}$, ${\tilde{c}_3}$, $\cdots$ in the curvature sector are all bare, and thus, they are not at the same footing as the loop-corrected constants in the effective QFT. They remain bare since matter loops have already been used up in forming the flat spacetime effective action $S_{eff}\left(\eta,\psi,\Lambda_\wp\right)$ in (\ref{eff-ac}), and there have remained thus no loops (quantum fluctuations) to induce or correct any interaction like (\ref{curv-sector-0}) or not. This means that the curvature sector parameters bear no sensitivity to $\Lambda_\wp$. This discord between the two sectors implies that the parameters ${\widetilde{G}}$, ${\tilde{c}_2}$, ${\tilde{c}_3}$, $\cdots$
are in fact all incalculable \cite{demir2016,demir2019}.
This problem, which reveals the difference between the classical field theories and the effective QFTs, can be overcome if the curvature sector in effective QFTs is also loop-corrected or loop-induced. Namely, curvature sector must arise from the effective QFT itself during the map of
$S_{eff}\left(\eta,\psi,\Lambda_\wp\right)$ into curved spacetime. This requirement falls outside the workings of the general covariance map in (\ref{covariance-map}) as it involves only the flat metric $\eta_{\mu\nu}$  in $S_{eff}\left(\eta,\psi,\Lambda_\wp\right)$. This means that it is necessary to construct a whole new transformation rule for taking effective QFTs into curved spacetime. In this regard, this first stage would be the determination of what parameters to transform other than the flat metric:  Field masses $m_i$? The UV cutoff $\Lambda_\wp$? Some other scale in  $S_{eff}\left(\eta,\psi,\Lambda_\wp\right)$?  The second stage would be the determination of the transformation rule itself (presumably some extension of the general covariance). The question of what to transform other than
$\eta_{\mu\nu}$ will be answered in Sec. \ref{sect:constantlog} below by revealing the symmetry properties of different UV sensitivities in $S_{eff}\left(\eta,\psi,\Lambda_\wp\right)$. The transformation rule itself, on the other hand, will be determined in Sec. \ref{sect:restore} by requiring that the transformation must be able to erase the anomalous gauge boson masses in (\ref{deltMV}).

\section{$\Lambda_\wp^2$ {\lowercase{vs.}} $\log\Lambda_\wp\,$: Symmetry Structures}
\label{sect:constantlog}
The UV cutoff $\Lambda_\wp$ is not just a mass scale. It is more than that. To see why, it suffices to scrutinize the two distinct roles it plays in shaping the effective QFTs: 
\begin{enumerate}[(1)]
\item The $\Lambda_\wp^2$ correction is additive as ensured by the scalar masses (\ref{deltmphi}). It breaks gauge symmetries explicitly as proven by the gauge boson masses (\ref{deltMV}). It breaks 
also Poincare (translation) symmetry \cite{wigner,wald} as it restricts loop momenta $\ell_\mu$ into the range $-\Lambda_\wp^2 \leq \eta_{\mu\nu} \ell^\mu \ell^\nu \leq \Lambda_\wp^2$.
Thus, $\Lambda_\wp^2$ has an affinity for spacetime curvature as both of  them break the Poincare symmetry. 

\item The $\log\Lambda_\wp$ correction is always multiplicative. It does not alter the symmetry structure of the quantity it multiplies.  The field masses $m_i$, for instance, respect both the gauge and  Poincare symmetries and so do the logarithmic corrections $\delta m^2_i \propto m^2_i \log \Lambda_\wp$. (The scalar masses in (\ref{deltmphi}) set an example with $m_\phi^2$ being the Casimir invariants of the Poincare group \cite{wigner}.)
The same is true for all the QFT couplings. In parallel with this, as follows from the item (1) above, $\Lambda_\wp^2$ breaks both the gauge and Poincare symmetries and so does the $\Lambda_\wp^2 \log\Lambda_\wp$ (which can arise at higher loops). It thus turns out that  $\log\Lambda_\wp$ respects both the gauge and Poincare symmetries. It can have therefore no affinity for  (Poincare-breaking) spacetime curvature.
\end{enumerate}
The two distinct roles played by $\Lambda_\wp$ are contrasted in Table \ref{table-3}. The main lesson is that $\log\Lambda_\wp$ must remain intact
under a possible correspondence between $\Lambda_\wp^2$ and curvature on the basis of their Poincare affinity.

\begin{table}
\caption{\label{table-3} The symmetry structures of $\Lambda_\wp^2$ and $\log\Lambda_\wp$, and their affinity to curvature. $\Lambda_\wp^2$ breaks both the gauge and Poincare symmetries but $\log\Lambda_\wp$ does not (as it always comes multiplicatively and  does not alter the symmetry of the term it multiplies).}
\begin{tabular}{|l|l|l|l|}
\hline
{}&Gauge Symmetry&Poincare Symmetry & Affinity to Curvature\\
\hline\hline
$\Lambda_\wp^2$ & ${\rm X}$  & ${\rm X}$ & $\checkmark$\\\hline
$\log\Lambda_\wp$ & $\checkmark$ &$\checkmark$ & ${\rm X}$\\
\hline
\end{tabular}
\end{table}

\section{Erasure of Anomalous Gauge Boson Masses by Emergent Curvature}
\label{sect:restore}
In view of the conclusion arrived in Sec. \ref{sect:hownot},  spacetime curvature must emerge from within $S_{eff}\left(\eta,\psi,\Lambda_\wp\right)$ in order not to hinge on arbitrary, incalculable constants as in (\ref{map-eff-ac}).
This condition can be taken to imply that there must exist some covariance relation between the mass scales in $S_{eff}\left(\eta,\psi,\Lambda_\wp\right)$
and spacetime curvature. To determine if there exists such a relation, it proves effectual to focus first on the gauge sector, whose   anomaly action (\ref{deltSV}) breaks gauge symmetries explicitly \cite{ccb}. It continues to break in curved spacetime if carried there via  (\ref{map-eff-ac}). But the gauge anomaly (\ref{deltSV}) is a pure $\Lambda_\wp^2$ effect and, in view of the Poincare affinity between $\Lambda_\wp^2$ and curvature (as revealed  in Sec. \ref{sect:constantlog}, Table \ref{table-3}), it is legitimate to ask a pivotal question: Is it possible to carry the effective QFT in (\ref{eff-ac}) into curved spacetime in a way erasing the anomalous gauge boson mass term (\ref{deltSV})? It will take a set of progressive steps to find out but the answer will turn out
to be  ``yes" \cite{demir2016,demir2019}:

{\bf Step 1.} The starting point of investigation is the self-evident identity 
\begin{eqnarray}
\delta S_V\left(\eta, \Lambda_\wp^2\right)= \delta S_V\left(\eta, \Lambda_\wp^2\right) - I_V(\eta) + I_V(\eta) 
\label{1st}
\end{eqnarray}
involving the gauge-invariant kinetic construct
\begin{eqnarray}
I_V(\eta) &=& \int d^{4}x \sqrt{-\eta}
\frac{c_V}{2} {\mbox{tr}}\!\left[ \eta_{\mu\alpha} \eta_{\nu\beta}V^{\mu\nu} V^{\alpha\beta}\right]\label{IV-first}\\ &=&
\int d^{4}x  \sqrt{-\eta} c_V {\mbox{tr}}\!\left[V^{\mu}\left( -D_{\mu\nu}^2 \right)V^{\nu} + {\partial}_{\mu} \left(\eta_{\alpha\beta} V^{\alpha} V^{\beta\mu}\right)\right]
\label{IV-second}
\end{eqnarray}
whose second line, obtained via by-parts integration of the first line,  consists of a surface term (the total divergence) and  inverse propagator $D_{\mu\nu}^2 = D^2 \eta_{\mu\nu} - D_{\mu}D_{\nu}-iV_{\mu\nu}$ with $D^2 = \eta^{\mu\nu} D_\mu D_\nu$ such that $D_\mu=\partial_\mu + iV_\mu$ is the gauge-covariant derivative, $V_\mu = V_\mu^a T^a$ is the gauge field with gauge group generators $T^a$, and $V_{\mu\nu}=V^a_{\mu\nu} T^a$ is the field strength tensor with the components $V^a_{\mu\nu} = \partial_\mu V^a_\nu - \partial_\nu V^a_\mu + i f^{a b c} V^b_\mu V^c_\nu$ ($f^{a b c}$ are structure constants). Now, the identity (\ref{1st}) can be put into the equivalent form
\begin{eqnarray}
\!\!\!\!\delta S_{V}\!\left(\eta, \Lambda_\wp^2\right) &\!=\!&
 - I_V(\eta) \nonumber\\&\!+\!&\!\int\!\! d^{4}x  \sqrt{-\eta} c_V {\mbox{tr}}\!\!\left[V^{\mu}\left( -D_{\mu\nu}^2 + \Lambda_\wp^2 \eta_{\mu\nu} \right)\!V^{\nu} + {\partial}_{\mu} \left(\eta_{\alpha\beta} V^{\alpha} V^{\beta\mu}\right)\right]
\label{2nd}
\end{eqnarray}
after, at the right-hand side of (\ref{1st}), $\delta S_V$ is replaced with its expression in (\ref{deltSV}), $``+I_V"$ is replaced with its expression in (\ref{IV-second}), and yet $``-I_V"$ is left untouched (kept as in (\ref{IV-first})).

{\bf Step 2.} Now, the rearranged gauge boson anomalous mass action in (\ref{2nd}) gets to curved spacetime via  the general covariance map (\ref{covariance-map}) to take there the ``curved" form  
\begin{eqnarray}
\!\!\!\!\!\delta S_{V}\!\left(g, \Lambda_\wp^2\right) &\!=\!&
 - I_V(g) \nonumber\\&\!+\!&\!\int\!\! d^{4}x  \sqrt{-g} c_V {\mbox{tr}}\!\!\left[V^{\mu}\left( -{\mathcal{D}}_{\mu\nu}^2 + \Lambda_\wp^2 g_{\mu\nu} \right)V^{\nu} + \nabla_{\mu} \left(g_{\alpha\beta} V^{\alpha} V^{\beta\mu}\right)\right]
\label{3rd}
\end{eqnarray}
in which  $\nabla_{\mu}$ is the spacetime covariant derivative with respect to the Levi-Civita connection (\ref{LC-conn}), ${\mathcal{D}}_{\mu}=\nabla_\mu + iV_\mu$ is the gauge-covariant derivative with respect to $\nabla_{\mu}$ so that ${\mathcal{D}}^2 = g^{\mu\nu} {\mathcal{D}}_\mu {\mathcal{D}}_\nu$ and ${\mathcal{D}}_{\mu\nu}^2 = {\mathcal{D}}^2 g_{\mu\nu} - {\mathcal{D}}_{\mu} {\mathcal{D}}_{\nu} - iV_{\mu\nu}$.

{\bf Step 3.} Now, a closer look at  the action (\ref{3rd}) reveals a crucial property:
\begin{enumerate}[(1)]
    \item if $\Lambda_\wp^2 g_{\mu\nu}$ were replaced with the Ricci curvature $R_{\mu\nu}\left({}^{g}\Gamma\right)$, and
    
    \item if $c_V$ (which can involve $\log\Lambda_\wp$ at higher loops) were held intact under (1)
\end{enumerate}
then ${{\delta S_{V}\left(g,\Lambda_\wp^2\right)}}$ would reduce to zero identically. To see this, one first 
replaces $\Lambda_\wp^2 g_{\mu\nu}$ with $R_{\mu\nu}\left({}^{g}\Gamma\right)$ in (\ref{3rd}) to get
\begin{eqnarray}
\!\!\!\!\!\!\delta S_{V}\!\left(g, R\right) &\!=\!&
 - I_V(g) \nonumber\\&\!+\!&\!\int\!\! d^{4}x  \sqrt{-g} c_V {\mbox{tr}}\!\!\left[V^{\mu}\!\left( -{\mathcal{D}}_{\mu\nu}^2 + R_{\mu\nu}\left({}^{g}\Gamma\right) \right)V^{\nu} + \nabla_{\mu}\! \left(g_{\alpha\beta} V^{\alpha} V^{\beta\mu}\right)\right]
\label{3rdp}
\end{eqnarray}
and then integrates (\ref{3rdp}) by parts using  $\left[{\mathcal{D}}_{\mu}, {\mathcal{D}}_{\nu}\right] = R_{\mu\nu}\left({}^{g}\Gamma\right) + iV_{\mu\nu}$ to arrive at 
\begin{eqnarray}
\!\!\!\!\delta S_{V}\!\!\left(g, \Lambda_\wp^2\right) \!=\!
 - I_V(g) + \!\!\int\!\! d^{4}x \sqrt{-g}
\frac{c_V}{2} {\mbox{tr}}\!\!\left[ g_{\mu\alpha} g_{\nu\beta}V^{\mu\nu} V^{\alpha\beta}\right] \!=\! -I_V(g) + I_V(g) 
\label{3rdpp}
\end{eqnarray}
which reduces to zero identically, as claimed above. This result holds  provided that $\Lambda_\wp^2 g_{\mu\nu}$ is replaced 
with $R_{\mu\nu}\left({}^{g}\Gamma\right)$ in (\ref{3rd}) and provided that this replacement leaves $c_V$ (in fact, $\log\Lambda_\wp$) intact. It is striking that the conditions (1) and (2) above are, respectively,  the first and the second rows of Table \ref{table-3} in Sec. \ref{sect:constantlog}.  This accord between the erasure of the anomalous gauge boson masses and the Poincare structure of the UV sensitivities of the QFT can be taken as a confirmation of the applied method.
 
It seems all fine. But actually there is a serious inconsistency problem here. Indeed, in the flat limit  ($g_{\mu\nu} \leadsto \eta_{\mu\nu}$) curvature remains nonzero ($R_{\mu\nu}\left({}^{g}\Gamma\right)\leadsto \Lambda_\wp^2 \eta_{\mu\nu}$). If it were not for this contradiction metamorphosis of $\Lambda_\wp^2 g_{\mu\nu}$ into curvature (confirmed by Table \ref{table-3}) would completely erase the anomalous gauge boson mass (\ref{deltSV}) and solve the CCB  \cite{demir2016,demir2019}. 

{\bf Step 4.} The inconsistency above can be remedied by introducing a more general map \cite{affine,demir2019}
\begin{eqnarray}
\label{emerge-curve-1}
\Lambda_\wp^2 g_{\mu\nu}\, \hookrightarrow\, {\mathbb{R}}_{\mu\nu}\left(\Gamma\right)
\end{eqnarray}
in which ${\mathbb{R}}_{\mu\nu}(\Gamma)$ is the Ricci curvature of a symmetric affine connection $\Gamma^{\lambda}_{\mu\nu}$.  (Here, $\Gamma^{\lambda}_{\mu\nu}$ 
and ${\mathbb{R}}_{\mu\nu}(\Gamma)$
have, respectively, nothing to do with ${}^g\Gamma^\lambda_{\mu\nu}$ in (\ref{LC-conn}) and $R_{\mu\nu}({}^g\Gamma)$ in (\ref{Ricci-met}), as shown contrastively in Table \ref{table-2}.) The metamorphosis of $\Lambda_\wp^2 g_{\mu\nu}$ into ${\mathbb{R}}_{\mu\nu}(\Gamma)$ goes parallel with the metamorphosis of $\eta_{\mu\nu}$ into $g_{\mu\nu}$ as a correspondence between physical quantities in the flat and curved spacetimes, and removes the inconsistency since 
the two maps, (\ref{covariance-map}) and (\ref{emerge-curve-1}), involve independent dynamical variables. In fact, affine curvature can well be the substance that fixes the vacuousity \cite{kretschmann} of general covariance. 
In view of this fixture, it proves efficacious to introduce the extended general covariance (EGC)
\begin{eqnarray}
\label{covariance}
\!\!\!\!\!\underbrace{\overbrace{S_{eff}\!\left(\partial\psi,\eta,\log\!\Lambda_\wp,\Lambda_\wp^2\!\right) \!\hookrightarrow\! S_{eff}\!\left(\nabla\psi, g,\log\!\Lambda_\wp,\Lambda_\wp^2\!\right)}^{\rm general\, covariance} \!\hookrightarrow\! S_{eff}\!\left(\nabla\psi,g,\log\Lambda_\wp,{\mathbb{R}}\!\right)}_{\rm extended\, general\, covariance\, (EGC)}
\end{eqnarray}
by combining the affine curvature map in (\ref{emerge-curve-1}) with the general covariance map in (\ref{covariance-map}) on the effective action $S_{eff}$ in (\ref{eff-ac}). The underlying symmetry
structure is given in Table \ref{table-3} in Sec. \ref{sect:constantlog}. The EGC reduces to the usual general covariance when $\Lambda_\wp$ is absent.

{\bf Step 5.} The EGC map in (\ref{covariance}) takes the action (\ref{3rd}) into 
\begin{eqnarray}
\!\!\!\!\!\!\delta S_{V}\!\left(g, {\mathbb{R}}\right) &\!=\!&
 - I_V(g) \nonumber\\&\!+\!&\!\int\!\! d^{4}x  \sqrt{-g} c_V {\mbox{tr}}\!\!\left[V^{\mu}\!\left( -{\mathcal{D}}_{\mu\nu}^2 + {\mathbb{R}}_{\mu\nu}\left(\Gamma\right) \right)V^{\nu} + \nabla_{\mu}\! \left(g_{\alpha\beta} V^{\alpha} V^{\beta\mu}\right)\right]
\label{tilded-3}
\end{eqnarray}
which reduces to 
\begin{eqnarray}
 {{\delta S_{V}\left(g, {\mathbb{R}}, R\right)}} = \int d^{4}x  \sqrt{-g} {c_V} {\mbox{tr}}\left[V^{\mu}\left( {\mathbb{R}}_{\mu\nu}\left(\Gamma\right) - R_{\mu\nu}\left({}^{g}\Gamma\right)\right) V^{\nu}\right]
\label{tilded-3p}
\end{eqnarray}
for the same reason that (\ref{3rdp}) reduced to (\ref{3rdpp}). This resultant action, which shows metamorphosis of the anomalous gauge boson masses in (\ref{deltSV}) into curvature terms,  is a truly metric-affine action \cite{affine,damianos} in that it involves both the affine and metrical curvatures (see Table \ref{table-2}). The ``$\delta S_V$" is no longer a gauge boson mass term. It is in this sense that the anamalous gauge boson masses get erased. The fate of the anomaly (the CCB), as will be analyzed below, is determined by the dynamics of the affine connection $\Gamma^{\lambda}_{\mu\nu}$.

\begin{table}
\caption{\label{table-2} Basic geometrical objects in flat and curved (metrical or affine) spacetimes.}
\begin{tabular}{|l|l|l|l|}
\hline
{}  &Flat &Curved (metrical) & Curved (affine)\\
\hline\hline
metric & $\eta$  & $g$ & \\\hline
connection & 0  & ${}^g\Gamma$ & $\Gamma$ \\\hline
covariant derivative & $\partial$ &$\nabla = \partial + {}^g\Gamma$ & ${}^{\Gamma}\nabla = \partial + \Gamma$\\\hline
curvature & 0 &$R({}^g\Gamma)$ & ${\mathbb{R}}(\Gamma)$\\
\hline
\end{tabular}
\end{table}

{\bf Step 6.} It is the curvature sector that decides on if  ${\mathbb{R}}_{\mu\nu}(\Gamma)$ comes close to  $R_{\mu\nu}\left({}^{g}\Gamma\right)$ to suppress the
gauge anomaly, that is, the action (\ref{tilded-3p}).  It originates from  (\ref{deltSOphi}) plus (\ref{deltSV}) by way of the  EGC in (\ref{covariance}), and takes the compact form 
\begin{eqnarray}
{\rm ``curvature\ sector"} &=& \int d^4x \sqrt{-g}\Bigg\{-{\mathbb{Q}}^{\mu\nu} {\mathbb{R}}_{\mu\nu}(\Gamma) + \frac{1}{16} c_\varnothing \left(g^{\mu\nu} {\mathbb{R}}_{\mu\nu}(\Gamma)\right)^2\nonumber\\ &-&  c_{V} {\mbox{tr}}\!\left[V_{\mu}V_{\nu}\right] R_{\mu\nu}({}^g\Gamma)\Bigg\}
\label{action-affine-2px}
\end{eqnarray}
after utilizing (\ref{tilded-3p}) and introducing 
\begin{eqnarray}
\label{q-tensor}
{\mathbb{Q}}_{\mu\nu} = \left(\frac{1}{4}\sum\limits_{i} c_{\psi_i} m_i^2 +  \frac{1}{4} c_\phi \phi^{\dagger} \phi + \frac{1}{8} c_\varnothing g^{\alpha\beta} {\mathbb{R}}_{\alpha\beta}(\Gamma)\right) g_{\mu\nu} - c_{V} {\mbox{tr}}\!\left[V_{\mu}V_{\nu}\right]
\end{eqnarray}
as a disformal metric \cite{disformal} typical of the metric-affine geometry \cite{affine,damianos,shimada}. (Transmutations of various objects from geometry to geometry are given in Table \ref{table-2}.) The metric-affine curvature sector (\ref{action-affine-2px}), which involves not a single incalculable constant, is precisely the structure anticipated at the end of Sec. \ref{sect:hownot} after realizing problems with the by-hand curvature sector in (\ref{curv-sector-0}). The EGC  in (\ref{covariance}) seems to have done the job.

{\bf Step 7.}  The affine dynamics,  which follows from the requirement that the curvature sector (\ref{action-affine-2px}) must remain stationary against variations in $\Gamma^{\lambda}_{\mu\nu}$, takes the compact form (affine covariant derivative ${}^\Gamma\nabla$ defined in Table \ref{table-2})
\begin{eqnarray}
\label{gamma-eom}
{}^{\Gamma}\nabla_{\lambda} {\mathbb{Q}}_{\mu\nu} = 0
\end{eqnarray}
after replacing the affine curvature \cite{affine,damianos}
\begin{eqnarray}
\label{affine-curv}
{\mathbb{R}}_{\mu\nu}\left(\Gamma\right) = \partial_\alpha \Gamma^\alpha_{\mu\nu} - \partial_\nu \Gamma^\alpha_{\alpha \mu} +  \Gamma^\beta_{\mu\nu}  \Gamma^\alpha_{\alpha\beta} - \Gamma^\beta_{\mu\alpha}  \Gamma^\alpha_{\nu\beta}
\end{eqnarray}
in the curvature sector in (\ref{action-affine-2px}). The solution of the equation of motion (\ref{gamma-eom})
\begin{eqnarray}
\label{gamma-gammag-2}
\Gamma^{\lambda}_{\mu\nu}={}^{g}\Gamma^{\lambda}_{\mu\nu} + \frac{1}{2} ({\mathbb{Q}}^{-1})^{\Lambda_\wp\rho} \left({{\nabla}}_{\mu} {\mathbb{Q}}_{\nu\rho} + {{\nabla}}_{\nu} {\mathbb{Q}}_{\rho\mu} - {{\nabla}}_{\rho} {\mathbb{Q}}_{\mu\nu}\right)
\end{eqnarray}
is a first order nonlinear partial differential equation for $\Gamma^\lambda_{\mu\nu}$ since ${\mathbb{Q}}_{\mu\nu}$ involves not only the scalars $\phi$ and vectors $V_\mu$ but also the affine curvature  ${\mathbb{R}}_{\mu\nu}(\Gamma)$ in (\ref{affine-curv}). This means that $\Gamma^{\lambda}_{\mu\nu}$ from (\ref{gamma-gammag-2}) can have degrees of freedom beyond $\phi$, $V_\mu$ and ${}^{g}\Gamma^{\lambda}_{\mu\nu}$. Nevertheless, a short glance at ${\mathbb{Q}}_{\mu\nu}$ reveals that it has in it the inverse Newton-Cavendish constant
\begin{eqnarray}
\label{G_N}
\frac{1}{G_N} = 4 \pi \left(\sum\limits_{i} c_{\psi_i} m_i^2 + c_\phi \langle \phi^{\dagger} \phi \rangle\right)
\end{eqnarray}
which is set by the masses $m_i$ of the QFT fields $\psi_i$ and vacuum expectation values $\langle \phi \rangle$ of the QFT scalars. It is the largest known mass scale (the Planck scale $M_{Pl} = (8\pi G_N)^{-1/2}$), and its enormity enables one to expand $({\mathbb{Q}}^{-1})^{\mu\nu}$ as 
\begin{eqnarray}
\label{inv-Qmunu}
({\mathbb{Q}}^{-1})^{\mu\nu} &=& 16 \pi G_N g^{\mu\nu} -  (16 \pi G_N)^2 \Big( \frac{c_\phi}{4} \left(\phi^{\dagger} \phi -\langle \phi^{\dagger} \phi\rangle\right) g^{\mu\nu}  + \frac{c_\varnothing}{8} g^{\alpha\beta} {\mathbb{R}}_{\alpha\beta}(\Gamma) g^{\mu\nu}\nonumber\\&-& c_{V} {\mbox{tr}}\!\left[V^{\mu}V^{\nu}\right]\Big) + {\mathcal{O}}\left(G_N^3\right)
\end{eqnarray}
so that $\Gamma^{\lambda}_{\mu\nu}$ in (\ref{gamma-gammag-2}) becomes
\begin{eqnarray}
\label{gamma-gammag-2-til}
\Gamma^{\lambda}_{\mu\nu}={}^{g}\Gamma^{\lambda}_{\mu\nu} + 8 \pi G_N \left({{\nabla}}_{\mu} {\mathbb{Q}}_{\nu\rho} + {{\nabla}}_{\nu} {\mathbb{Q}}_{\rho\mu}  {{\nabla}}_{\rho} {\mathbb{Q}}_{\mu\nu}\right) + {\mathcal{O}}\left(G_N^2\right)
\end{eqnarray}
and ${\mathbb{R}}_{\mu\nu}\left(\Gamma\right)$ in (\ref{affine-curv}) takes the form
\begin{eqnarray}
{\mathbb{R}}_{\mu\nu}(\Gamma) = R_{\mu\nu}({}^{g}\Gamma) + 4 \pi G_N \left(\nabla^2\right)_{\mu\nu}^{\alpha\beta} {\mathbb{Q}}_{\alpha\beta} +  {\mathcal{O}}\left(G_N^2\right)
\label{expand-curv}
\end{eqnarray}
where   $\left(\nabla^2\right)_{\mu\nu}^{\alpha\beta} = \nabla^{\alpha}\, \nabla_{\mu} \delta^{\beta}_{\nu} + \nabla^{\beta}\, \nabla_{\mu} \delta^{\alpha}_{\nu} - \Box \delta^{\alpha}_{\mu} \delta^{\beta}_{\nu} - \nabla_{\mu}\, \nabla_{\nu} g^{\alpha\beta} + (\mu \leftrightarrow \nu)$ with $\Box=g^{\mu\nu}\nabla_\mu \nabla_\nu$.   In the expansions  (\ref{gamma-gammag-2-til}) and (\ref{expand-curv}),  iteration of ${\mathbb{R}}_{\mu\nu}(\Gamma)$ 
order by order in $G_N$ reveals that dependencies on  ${\mathbb{R}}_{\mu\nu}(\Gamma)$ reside always in the remainder (one higher-order in $G_N$). This means that $\Gamma^{\lambda}_{\mu\nu}$ and ${\mathbb{R}}_{\mu\nu}(\Gamma)$ get effectively integrated out of the dynamics to leave behind only the scalars $\phi$, vectors $V_\mu$ and the Levi-Civita connection ${}^{g}\Gamma^{\lambda}_{\mu\nu}$. This solution of ${\mathbb{R}}_{\mu\nu}(\Gamma)$ causes the action (\ref{tilded-3p}) to vanish
\begin{eqnarray}
{\delta S_{V}\!\left(g, {\mathbb{R}}, R\right)} &=& \int d^{4}x  \sqrt{-g} {c_V} {\mbox{tr}}\!\left[V^{\mu}\left( {\mathbb{R}}_{\mu\nu}\left(\Gamma\right) - R_{\mu\nu}\left({}^{g}\Gamma\right)\right)V^{\nu}\right]\nonumber\\ &=& \int d^4x \sqrt{-g} \left\{0 + {\mathcal{O}}\!\left(G_N\right)\right\}
\label{gauge-reduce}
\end{eqnarray}
up to an anomaly-plagued ${\mathcal{O}}\!\left(G_N\right)$ remainder
\begin{eqnarray}
\int d^4x \sqrt{-g}\, {\mathcal{O}}\!\left(G_N\right) \!= \!\! \int\!\! d^{4}x  \sqrt{-g}  \left\{4 \pi G_N  {\mbox{tr}}\!\left[Q^{\mu\nu} \left(\nabla^2\right)_{\mu\nu}^{\alpha\beta} {\mathbb{Q}}_{\alpha\beta}\right] + {\mathcal{O}}\!\left(G_N^2\right)\right\}
\label{gauge-reduce-next}
\end{eqnarray}
which is an all-order derivative interaction following from (\ref{inv-Qmunu}). It never generates any mass for  $\phi$ and $V_\mu$. In effect, anomalous gauge boson  masses have been completely erased. The remnant anomaly effects in (\ref{gauge-reduce-next}), which go like $G_N E^2$ at boson-boson (say, photon-photon, Higgs-gluon,  $\dots$) collisions of energy $E$, are too tiny to leave any detectable signatures at current collider experiments. They become important though at scales near $G_N^{-1/2}$ as then the power series expansion in (\ref{inv-Qmunu}) fails, the affine connection remains unintegrated-out, and the fifth-force type effects start coming into play \cite{will,5th}.

In summary, the EGC (\ref{covariance})  has converted the anomalous gauge boson mass term (\ref{deltSV}) into the metric-affine action (\ref{tilded-3p}), and integration of the affine curvature out of the dynamics has killed  (\ref{tilded-3p}) to leave behind only the doubly Planck-suppressed CCB-plagued remainder in (\ref{gauge-reduce-next}). The key determinant is the EGC. It has enabled curvature to  {\it symmerge}, that is, emerge in a way restoring gauge symmetries (albeit with an ${\mathcal{O}}\left(G_N\right)$ tiny breaking \cite{anomaly,ccb}).

\section{QFT-GR Concord}
\label{sect:concord}
It is now time to carry the entire flat spacetime effective QFT in (\ref{eff-ac}) into curved spacetime. To this end, the EGC in (\ref{covariance}), which has arrived on the scene as if a deus ex machina, provides the requisite transformation rules (in parallel with Table \ref{table-3}).
   
The power-law part of the flat spacetime effective action in (\ref{eff-ac}), which has turned into
the metric-affine curvature sector in (\ref{action-affine-2px}) via the EGC in (\ref{covariance}), leads to the GR action
\begin{eqnarray}
S_{\rm GR}(g,\phi) &=& \int d^4x \sqrt{-g}\Bigg\{-\frac{R(g)}{16\pi G_N} -  \frac{c_\phi }{4} \left(\phi^\dagger \phi - \langle \phi^\dagger \phi\rangle\right)R(g) - \frac{c_\varnothing}{16}  R^2(g)\nonumber\\ &+& {\mathcal{O}}\left(G_N\right)\Bigg\}
\label{action-GR}
\end{eqnarray}
after integrating out ${\mathbb{R}}_{\mu\nu}(\Gamma)$ from (\ref{action-affine-2px}) via its  solution in (\ref{expand-curv}). Each and every coupling ($G_N, c_\phi, c_\varnothing, \dots$) in this action is a bona fide quantum effect. In fact, the UV sensitivity problems revealed in Sec. \ref{sect:problems} are seen to have all disappeared en route to (\ref{action-GR}). They have disappeared through the EGC as follows:  
\begin{enumerate}[(a)]
 \item EGC suppressed the CCB \cite{anomaly,demir2016,ccb}  by mapping the anomalous gauge boson masses in  (\ref{deltMV}) into the ${\mathcal{O}}\left(G_N\right)$ derivative interactions in (\ref{gauge-reduce-next}),
 
\item EGC eliminated the big hierarchy problem \cite{veltman} by converting the $\Lambda_\wp^2$ part of the scalar masses in (\ref{deltmphi}) into non-minimal coupling $c_\phi/4$ between the scalars $\phi$ and the curvature scalar $R(g)$ \cite{non-min} (This does not mean that the hierarchy problem is solved because  logarithmic corrections in (\ref{deltmphi}) are not dealt with yet.), and 
   
\item EGC prevented occurrence of the CCP \cite{ccp,ccp2} by transfiguring the $\Lambda_\wp^2$ and $\Lambda_\wp^4$ parts of the vacuum energy (\ref{deltV}) into the Einstein-Hilbert and quadratic curvature terms \cite{fR}, respectively (This does not mean that the CCP is solved because logarithmic corrections in (\ref{deltV}) are not dealt with yet.). 
\end{enumerate}
The resolution of these notorious power-law UV sensitivity problems is one important difference between symmergent gravity and Sakharov's induced gravity \cite{sakharov,visser}.

The logarithmic part $\delta S_{l}\left(\eta,\psi,\log\Lambda_\wp\right)$ plus the classical part $S_{c}\left(\eta,\psi\right)$ of the flat spacetime effective action (\ref{eff-ac}) gets to curved spacetime 
\begin{eqnarray}
\label{action-reg}
S_{\rm QFT}\left(g,\psi,\log\Lambda_\wp\right) = S_{c}\left(g,\psi\right) + \delta S_{l}\left(g,\psi,\log\Lambda_\wp\right) 
\end{eqnarray} 
since $\log\Lambda_\wp$ remains unchanged under the EGC map in (\ref{covariance}). In essence, the logarithmic UV sensitivity is equivalent to the dimensional regularization \cite{cutoff-dimreg}. Indeed, the formal equivalence $\log \Lambda_\wp^2 \equiv {1}/{\epsilon} -\gamma_E + 1 + \log 4\pi \mu^2$ can always be used to translate (\ref{action-reg}) into dimensional regularization scheme in $4+\epsilon$ dimensions with  Euler-Mascheroni constant $\gamma_E \approx 0.577$ and  renormalization scale $\mu$. The removal of the $1/\epsilon$ pieces, for instance, corresponds to minimal subtraction scheme renormalization of  (\ref{action-reg}) \cite{dim-reg2,dim-reg3}.

The EGC images in  (\ref{action-GR}) and (\ref{action-reg}) of the  flat spacetime effective action in (\ref{eff-ac}) combine to form the intertwined whole 
\begin{eqnarray}
S_{{\rm QFT \cup GR}} = S_{\rm QFT}(g, \psi, \log\Lambda_\wp)  + S_{\rm GR}(g,\phi)
\label{concord}
\end{eqnarray}
which describes matter by the regularized QFT and geometry by the GR. It is the QFT-GR concord sought for. Its primary features are as follows: 
\begin{enumerate}
\item The fact that a QFT-GR concord is formed can be understood via scattering amplitudes. To this end, as an illustration, it can prove useful to take a glance at $f_1 f_2 \rightarrow \phi_1 \phi_2 \phi_3$ scattering -- coannihilation of two fermions $f_1$ and $f_2$ into three scalars $\phi_1$, $\phi_2$, $\phi_3$.  It rests on the field operator structure
\begin{eqnarray} 
\int d^4x \sqrt{-g(x)} \int d^4y  \sqrt{-g(y)}\, {\overline{f}}_1(x) f_2(x) \Bigg\{h_{f_1 f_2 \phi }   \Delta_\phi(x,y) \lambda_{\phi \phi_1 \phi_2 \phi_3} \nonumber\\ +  \frac{{\widetilde{\lambda}}_{f_1 f_2 \phi_1 \phi_2 \phi_3}}{\Lambda_{QFT}^2} \delta^{(4)}(x,y)   \Bigg\}{\phi}_1(y) {\phi}^\dagger_2(y) {\phi}_3(y)
\label{operators}
\end{eqnarray}
which has a non-contact part mediated by the scalar propagator $\Delta_\phi$, and a contact part suppressed by the QFT scale $\Lambda_{QFT}$ (and hence
left out of the effective action (\ref{eff-ac})).  The Yukawa coupling $h$, the quartic coupling $\lambda$, and the scalar mass $m_\phi$ in $\Delta_\phi$ are not tree-level objects but rather loop-level objects corrected by the flat spacetime loops. Likewise, the higher-order coupling ${\widetilde{\lambda}}$ originates from the flat spacetime loops. The field operators (like $\phi_1$ and $f_2$), on the other hand, are actually mean fields (${\bar {\phi}}_1$ and ${\bar f}_2$) averaged over their quantum fluctuations while forming the flat spacetime effective action in (\ref{eff-ac}). In symmergence, therefore, operator structures like (\ref{operators}) attain the mean-field form
\begin{eqnarray} 
\int d^4x \sqrt{-g(x)} \int d^4y  \sqrt{-g(y)}\, {\overline{\bar{f}}}_1(x) {\bar f}_2(x) \Bigg\{h_{f_1 f_2 \phi }   \Delta_\phi(x,y) \lambda_{\phi \phi_1 \phi_2 \phi_3} \nonumber\\+  \frac{{\widetilde{\lambda}}_{f_1 f_2 \phi_1 \phi_2 \phi_3}}{\Lambda_{QFT}^2} \delta^{(4)}(x,y)   \Bigg\}{\bar \phi}_1(y) {\bar \phi}^\dagger_2(y) {\bar \phi}_3(y)
\label{mean-field}
\end{eqnarray}
whose fields (like ${\tilde {\phi}}_1$ and ${\tilde f}_2$) are solutions of the associated wave equations in the curved spacetime of the metric $g_{\mu\nu}$. This means that scattering and decay rates can be analyzed as interactions of the relativistic wavefunctions in curved spacetime \cite{cl-scat-th}. In symmergence, therefore, QFT-GR concord is achieved not by quantizing gravity but by grounding on the flat spacetime effective QFT \cite{demir2016,demir2019}.

In the traditional approach of curved spacetime QFTs \cite{qft-cuvred}, attempts to determine scattering amplitudes  get stuck already at the operator stage in (\ref{operators})
simply because it is not possible to construct the $|{\rm in}\rangle$ and $|{\rm out}\rangle$ Fock states  \cite{wald,ashtekar}. Symmergence is immune to all these problems simply because it is based on flat spacetime effective QFTs.

\item The coupling $c_\phi/4$ in the GR action $S_{\rm GR}(g,\phi)$ is a loop factor expected to be a few $\%$ ($1.3\%$ in the SM). It
couples scalar curvature $R(g)$ to scalar fields $\phi$, and gives cause thus to Newton-Cavendish constant to vary with $\phi$ \cite{non-min}.  For high field 
values near $M_{GUT}\sim 10^{-2} M_{Pl}$, $G_N$ gets rescaled to $G_N (1-10^{-4} c_\phi) \sim G_N (1-10^{-6})$ 
since $\phi^\dagger \phi - \langle \phi^\dagger \phi\rangle\sim M_{GUT}^2$. This variation in $G_N$ remains well within the 
experimental uncertainty in $G_N$  \cite{codata}. The agreement with data gets better and better with lower and lower field values.  For higher field values, however,
 variation in $G_N$ exceeds the experimental range (by three orders of magnitude at $|\phi|\sim M_{Pl}$). This means that
$|\phi_{max}| \sim M_{GUT}$ is the largest field swing allowed. 

\item The Newton-Cavendish constant in $S_{\rm GR}(g,\phi)$
\begin{eqnarray}
\label{G_N-1loop}
\frac{1}{G_N} = 4 \pi \left(\sum\limits_{i} c_{\psi_i} m_i^2 + c_\phi \langle \phi^{\dagger} \phi \rangle\right)  \xrightarrow{\rm 1-loop} \frac{{\rm str}[m^2]}{8 \pi}  +  4 \pi c^{(1)}_\phi \langle \phi^{\dagger} \phi \rangle
\end{eqnarray}
must agree with empirical data \cite{will} for gravity to symmerge correctly.  It constrains  the QFT mass spectrum  as ${\rm str}\left[m^2\right]\sim M_{Pl}^2$ (barring  flat directions in which $\langle \phi \rangle \gg m_\phi$). Its
one-loop form reveals that the QFT particle spectrum must be dominated by bosons either in number or in mass or in both. It thus turns out that the Newton-Cavendish constant can be correctly induced in a QFT having 
\begin{enumerate}[(i)]
    \item either a light spectrum with numerous more bosons than fermions (for instance, $m_b \sim m_f \sim {\rm TeV}$ with $n_b-n_f \sim 10^{32}$),
    
    \item or a heavy spectrum with few more bosons than fermions (for instance, $m_b \sim m_f \lesssim M_{Pl}$ with $n_b-n_f \gtrsim 5$),
    \item or a sparse spectrum with net boson dominance.
\end{enumerate}
The QFT is best exemplified by the SM, which is fully confirmed by the LHC experiments and their priors.  Its spectrum yields $G_N \sim - ({\rm TeV})^{-2}$, 
which is unphysical in both sign and size. It is because of this inadequacy of the SM spectrum that symmergence predicts existence of new physics beyond the SM (BSM).  The BSM, whose
spectrum adds to (\ref{G_N-1loop}) to correct the SM result,  has no obligation to couple to the SM in a specific scheme and strength. It can thus form a completely decoupled black
sector \cite{demir2019,black} or a weakly coupled dark sector \cite{demir2019,darksector}, with distinctive
signatures for collider searches \cite{keremle}, dark matter searches  \cite{cemle}, and other  possible phenomena \cite{demir2019}.

\item The quadratic curvature term in the GR action $S_{\rm GR}(g,\phi)$ can facilitate 
Starobinsky inflation  \cite{starobinsky,Planck,irfan} since (at one loop)
\begin{eqnarray}
c_\varnothing = -\frac{(n_b-n_f)}{128 \pi^2}
\label{c-bos}
\end{eqnarray}
acquires right sign and size 
for $n_b-n_f \approx 10^{13}$. In case this constraint is not met, inflation can also be realized with the scalar fields $\phi$ in the spectrum \cite{Higgs-inflation,bauer}.

\item The vacuum energy in $S_{\rm QFT}(g, \psi, \log\Lambda_\wp)$
\begin{eqnarray}
V(\Lambda_\wp) &&= V\left(\langle\phi\rangle\right) + \sum_i {c}^{(l)}_{\varnothing\psi_i} m_i^4  \log \frac{m_{i}^2}{\Lambda_\wp^2} \nonumber\\ &&\xrightarrow{\rm 1-loop} V^{(1)}\left(\langle\phi\rangle\right) + 
\frac{1}{64
\pi^2}{\rm str}\left[{m^{4}}\log \frac{m^{2}}{\Lambda_\wp^2}\right]
\label{vac-en}
\end{eqnarray}
gathers together the relevant $\log\Lambda_\wp$ corrections in (\ref{deltSlog}) in the minimum of the scalar potential $V(\phi)$ at $\phi=\langle \phi\rangle$. Its
empirical value is $V_{emp} = \left(2.57\times 10^{-3}\ {\rm eV}\right)^4$ \cite{ccp2}, and the QFT vacuum must reproduce this specific 
value.  This constraint puts severe restrictions on the UV cutoff and other parameters of the QFTs.  This is what the CCP \cite{ccp} is all about. 
In Sakharov's induced gravity \cite{sakharov,visser}, for instance, 
induction of the Newton-Cavendish constant  fixes $\Lambda_\wp$ to a Planckian value, and this fix leads to an
${\mathcal{O}}\left(M_{Pl}^4\right)$ vacuum energy. 
In symmergence, however, $\log\Lambda_\wp$ is not fixed (as in Table \ref{table-3}), and it
can be fixed in a way suppressing the vacuum energy. 
Indeed, in view mainly of the one-loop value in (\ref{vac-en}), the rough concordance
\begin{eqnarray}
\Lambda_\wp^2 \sim {\rm str}\left[m^2\right]\sim M_{Pl}^2
\label{relation}
\end{eqnarray}
suppresses the logarithms and induces the Newton-Cavendish constant consistently. (This becomes clear especially when ${\rm str}\left[m^2\right]$ is saturated by one large mass.) It involves a
severe fine-tuning, and thus, it certainly  is not a solution to the CCP.  But it might be sign  of an underlying symmetry
principle or a dynamical theory of the Poincare breaking scale $\Lambda_\wp$. Symmergence cannot go beyond (\ref{relation}). (Clearly, the CCP can be approached by other methods like degravitation mechanisms \cite{ccp3,ccp4}.)

\item Light scalars $\phi_L$ in $S_{\rm QFT}(g, \psi, \log\Lambda_\wp)$ have their masses shifted as
\begin{eqnarray}
\label{deltmphi-lhp}
\delta m_{\phi_L}^2 = {c}^{(l)}_{ \phi_L \psi_H} m_{\psi_H}^2 \log \frac{m_{\psi_H}^2}{\Lambda_\wp^2} 
\end{eqnarray}
via their couplings ${c}^{(l)}_{ \phi_L \psi_H}$  to heavy fields $\psi_H$, as defined in (\ref{deltmphi}) as well as (\ref{deltSlog}).  
These logarithmic corrections are of a new kind because they are sensitive to  field masses not to the UV cutoff.  And they are crucial because heavier the $\psi_H$ larger 
the $\delta m_{\phi_L}^2$ and stronger the destabilization of the light sector of the QFT. Symmergence cannot solve this problem (EGC in (\ref{covariance}) leaves $\log\Lambda_\wp$ untouched)
but provides a viable way to avoid it. The thing is that induction of the Newton-Cavendish constant in (\ref{G_N-1loop}) 
is the only constraint on the QFT field spectrum and it does not require any special coupling scheme or strength among the fields. Namely, induction of the Newton-Cavendish constant is immune to with what strengths the QFT fields are coupled. This immunity enables QFTs to maintain their two-scale structure at the loop level by having their heavy and light sectors coupled in a way keeping them scale-split.  The implied coupling 
\begin{eqnarray}
\label{seesawic}
\left|{c}^{(l)}_{ \phi_L \psi_H}\right| \lesssim \frac{m_{\phi_L}^2}{m_{\psi_H}^2}  
\end{eqnarray}
is see-sawish in nature and capable of stabilizing the light sector, as ensured by (\ref{deltmphi-lhp}). This implies that only those QFTs having see-sawish couplings can withstand loop-induced heavy-light mixing.  And symmergence allows the see-sawish couplings \cite{demir2019}.

The importance of the see-sawish couplings is best revealed by examining the realistic case of the SM. The SM needs be extended for various empirical and conceptual reasons \cite{beyond}. The extension, a BSM sector, is formed by superpartners in supersymmetry, Kaluza-Klein modes in extra dimensions, and technifermions in technicolor. Each of these BSM sectors couples to the SM with the SM couplings themselves as otherwise their underlying symmetries get broken.  And this means that they never go to the see-sawish regime in (\ref{seesawic}) and, as a result, the SM Higgs sector gets destabilized by these heavy BSM sectors. Indeed, even the Planck-scale supersymmetry (generating no quadratic correction by its nature and merging with the SM through an intermediate-scale singlet sector) destabilizes the Higgs boson mass as in (\ref{deltmphi-lhp}) \cite{quiros}. It is for this non-see-sawish nature of theirs that all these BSM sectors have already been sidelined by the LHC experiments for certain mass ranges.

The BSM sector of the symmergence, required by  the Newton-Cavendish constant in (\ref{G_N-1loop}), differs from superpartners, Kaluza-Klein levels and technifermions by its congruence to the see-sawish couplings in (\ref{seesawic}). It is different in that it contains only those fields which enjoy the see-sawish regime, and such fields may conveniently be termed as {\it symmergeons}. In fact, new physics searches beyond the TeV domain must assume a BSM sector that does not destabilize the SM Higgs sector and, in this respect, possible discoveries at future experiments \cite{collider1} may fit to the symmergeons. 
\end{enumerate}
These salient points feature the field-theoretic and gravitational aspects of the QFT-GR concord in (\ref{concord}) in relation to its
cosmological \cite{de,darksector,cos-coll}, astrophysical \cite{cdm,darksector}, and  collider \cite{collider1,collider2} implications. 

\section{Conclusion and Future Prospects}
\label{sect:conc}
In confirmation of the title, emergent gravity has erased anomalous gauge boson masses as in (\ref{gauge-reduce}) and a QFT-GR concord is formed as in (\ref{concord}). The gauge anomaly (the CCB) is not completely  banished as it survives in doubly-Planck suppressed derivative interactions in (\ref{gauge-reduce-next}). The resultant QFT-GR concord describes matter by  (dimensionally) regularized QFT and geometry by the GR. In the realistic case of the SM, it predicts the existence of a BSM sector, which does not have to couple to the SM. It can form therefore a  weakly-interacting or completely non-interacting sector, and in either case it can give cause for various  cosmological, astrophysical and collider phenomena. It is with advancements in energy, intensity and cosmology frontiers that the QFT-GR concord will take shape, with the determination of its QFT part, for instance. 

The present work is a small step. It needs be furthered and deepened in various aspects:
\begin{itemize}
\item The first aspect  concerns the covariance between (UV cutoff)$^2$ and curvature  (based on the affinity structure in Table \ref{table-3}). The question is this: Can this covariance be given a more fundamental structure? The answer, which is far from obvious, may involve gauge-theoretic approach to metric-affine gravity \cite{hehl} or even to the affine gravity \cite{eddington1,eddington2} (in view of its quantization properties \cite{quant-grav}). The gauge-theoretic approach \cite{hehl,hehl2} may necessitate metric-independent structures like the Ehressmann connection \cite{ehresmann} (as well as the Finsler geometry \cite{finsler}). The gauge-theoretic (or some other) substructure can promote symmergence to a more fundamental status. 

\item The second aspect concerns implications for the QFT spectrum of the simultaneous realization of the Newton-Cavendish constant in (\ref{G_N-1loop}) and the Starobinsky inflation via (\ref{c-bos}). In fact, they seem to disagree on $n_b-n_f$ in the degenerate cases, and a sparse spectrum seems more plausible.  Needless to say, a detailed knowledge of the spectrum 
can help reveal symmetries for alleviating the CCP and generating the see-sawish couplings.

\item The third aspect is about the CCP \cite{ccp,ccp2}. A useful feature of symmergence is its leaving of $\log\Lambda_\wp$ free. Its use for cancellation of the vacuum energy is insightful but incomplete in that it is necessary to find an all-loop selection rule or symmetry to prevent the enormous fine-tuning involved. To this end, cancellations of the known components like the QCD and electroweak vacuum energies by the BSM contributions can be helpful in revealing the aforementioned symmetry structure. The envisaged symmetry (which might be inspired by mirror symmetry \cite{twin}) must correlate the SM and the BSM fields along with the freedom provided by $\log\Lambda_\wp$.

\item The fourth aspect refers to the see-sawish couplings of the symmergeons. This is about not the UV boundary but the inner structure of the QFT.  There is in general no known symmetry principle that can lead to see-sawish structure. The problem becomes clear especially in multi-scalar theories \cite{haber} and, in this regard, the mass-degeneracy-driven unification proposed in \cite{cemle} (see also \cite{keremle}) seems to be one likely approach. It is, however, more of a condition rather than a symmetry principle, and needs therefore be furthered (by implementing perhaps mass-sensitive extensions of the symmetries of the multi-Higgs doublet models \cite{2hdm}).

\item The fifth aspect is related to the astrophysical and cosmological implications of the QFT-GR concord. Indeed, in the symmergent GR in which the Newton-Cavendish constant is set by  (\ref{G_N-1loop}), the quadratic curvature term by (\ref{c-bos}), and the vacuum energy by (\ref{vac-en}) the cosmological and astrophysical environments can put strong constraints on the QFT spectrum. One of them is cosmic evolution and its implications in view mainly of the persisting Hubble tension \cite{hubble}.  Another of them is the dense media like neutron stars \cite{ns}. The solutions of the Einstein field equations in such environments, with the added feature that all (matter) couplings are already loop-corrected, can give information about the spectrum and the loop structure.

\item The sixth point concerns high curvature limit. Indeed, if the affine curvature takes Planckian  values (${\mathbb{R}}\sim 1/G_N$) then the expansion in (\ref{inv-Qmunu}) fails and the solution of the affine connection in (\ref{gamma-gammag-2-til})  breaks down. This means that its exact solution in (\ref{gamma-gammag-2}), which  is a first order non-linear partial differential equation for itself, will contain extra geometrical degrees of freedom not found in the Levi-Civita connection. These new geometrodynamical fields will couple to matter and contribute to the gauge anomaly, though anomalous gauge boson masses will still exactly vanish  since ${\mathbb{R}}_{\mu\nu}(\Gamma)$ in (\ref{affine-curv}) has always an ${R}_{\mu\nu}({}^g\Gamma)$ part that cancels out the ${R}_{\mu\nu}({}^g\Gamma)$ in (\ref{tilded-3p}). This dynamical picture shows that symmergence may lead to novel phenomena in high-curvature regions like the black holes \cite{hawking,bh}.
\end{itemize}
\noindent It is with the investigation of these six salient aspects plus various other collateral ones that the true potential of the symmergence will be revealed. 

\section*{Acknowledgements}
This work is supported in part by the T{\"U}B{\.I}TAK grant 118F387.

\end{document}